\documentclass[3p]{elsarticle}

\usepackage{lineno,hyperref}
\usepackage{bm,amsmath}

\journal{Computational and Theoretical Chemistry}

\newcommand{\bra}{\left<}
\newcommand{\ket}{\right>}

\newcommand{\kket}[1]{\left|#1\right>}

\newcommand{\ovl}[2]{\bra #1|#2\ket}
\newcommand{\av}[1]{\bra #1\ket}
\newcommand{\bT}{\mathbf{T}}
\newcommand{\tkj}{T_{k\to j}}

\begin{document}

\begin{frontmatter}

\title{Excited state dynamics of some nonsteroidal anti-inflammatory drugs: a surface-hopping
investigation.}

\author[madrid,london,pisa]{Neus Aguilera-Porta}
\ead{neus.aguilera@uam.es}

\author[madrid,madrid2]{In\'es Corral\corref{cor1}}
\ead{ines.corral@uam.es}

\author[london]{Jordi Munoz-Muriedas} 
\ead{jordi.4.munoz-muriedas@gsk.com}

\author[pisa]{Giovanni Granucci\corref{cor1}}
\ead{giovanni.granucci@unipi.it}

\cortext[cor1]{Corresponding author}

\address[madrid]{Departamento de Qu\'{\i}mica, Universidad Aut\'onoma de Madrid,
                 28049 Cantoblanco, Madrid, Spain}
\address[madrid2]{IADCHEM. Institute for Advanced Research in Chemistry, Universidad Aut\'onoma de Madrid,
                 28049 Cantoblanco, Madrid, Spain}              
\address[london]{Computational Toxicology, GlaxoSmithKline, Ware, Hertfordshire, UK}
\address[pisa]{Dipartimento di Chimica e Chimica Industriale, Universit\`a di Pisa,
               via Moruzzi 13, 56124 Pisa, Italy}

\begin{abstract}
  Here we present a computational investigation of the excited state dynamics of 5 different active medical substances (aspirin,
  ibuprofen, carprofen, suprofen, indomethacin) which belong to the family of nonsteroidal
  anti-inflammatory drugs (NSAIDs). The nonadiabatic dynamics simulations were performed using the
  surface hopping method, with electronic energies and couplings evaluated on the fly in a
  semiempirical framework. For aspirin, the solvent decay dynamics was also considered by inserting
  it in a cluster of water molecules, following a QM/MM scheme. A quite diverse behavior was observed for the
  systems considered, going from fast deactivation to the $S_0$ (aspirin and ibuprofen), to ultrafast
  intersystem crossing to the triplet manifold (carprofen and suprofen), or to the generation of long lived $S_1$
  states (indomethacin). To our knowledge, this is the first computational study of the photodynamics
  of these NSAIDs.
\end{abstract}


\end{frontmatter}


\section{Introduction}
Drugs are exposed to natural or artificial light along the pharmaceutical chain, from their
manufacture until their dispensation or even after administration. The interaction between a drug and 
ultraviolet (UV) light may give rise to phototoxic and/or photoallergic side effects.
Both kind of processes are referred to as ``photosensitivity'' \cite{Vallet2012,Musa2010}.
A very well known family of drugs which may give rise to photosensitive reactions 
is that of nonsteroidal anti-inflammatory drugs
(NSAIDs), which are used mainly as analgesic, antipyretic and anti-inflammatory agents
\cite{Vallet2012,Musa2010}.

Photosensitive reactions are complex processes, involving different pathways such as energy,
electron or hydrogen transfer, photodecomposition and photobinding. The investigation of the
photophysics of drugs or the photoreactive paths that might be activated upon photon absorption is
crucial for identifying photoinitiated side effects, as well as in the design of novel and improved drugs. 
In the literature, there are a few experimental and computational studies on the photophysics of NSAIDs,
most often based on steady state results
\cite{Florey1979,Budac1992,Vione2011,Bosca1997,Sheu2003,Du2014,Maity2015,Dabestani1993,Moore1988,
Ghatak2008,Weedon1991,Miotke2017}, 
or on quantum chemical calculations on the energetics of their excited states, mainly at DFT/TD-DFT level of theory 
\cite{Musa2010,Govindasamy2014,Ye2016,Musa2011,Musa2007}. 
However, investigations of the excited state dynamics of these molecules are still lacking.

In the present work, we considered five NSAIDs: aspirin (ASP), ibuprofen (IBU), carprofen (CAR),
suprofen (SUP) and indomethacin (IDM), see Figure \ref{fig:molecs}. Our aim is the study of their
excited state dynamics through nonadiabatic molecular dynamics simulations, in order to gain
information about their gas phase photophysics (singlet lifetimes, triplet quantum yields, and the
related mechanisms and pathways). In the case of ASP, the effect of the solvent (water) along the
relaxation dynamics has been simulated as well.  We believe that our results will be also helpful in
the selection of descriptors for quantitative structure-activity relationships (QSAR) models
\cite{Gubskaya2010} that could be used to predict photoactivity of new or untested derivatives of
these NSAIDs.

The outline of this paper is as follows. Section 2 introduces the computational
strategy adopted for the on the fly molecular dynamics simulations (a
methodological improvement is reported in the appendix), section 3
describes the excited states dynamics, and concluding remarks are offered in section 4.


\section{Method}
\begin{figure}
  \begin{center}
  \includegraphics[width=10.0cm]{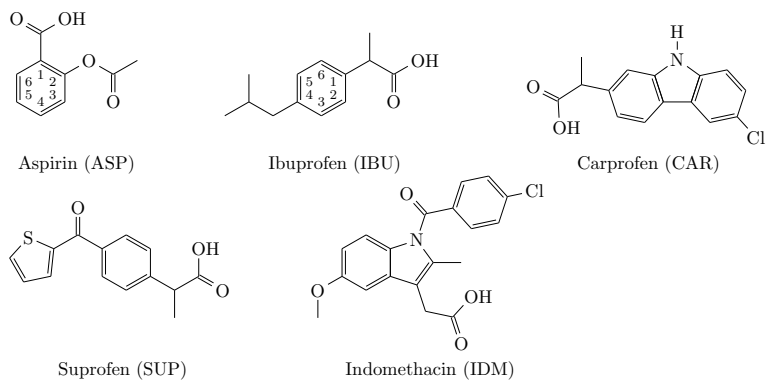}
  \caption{The nonsteroidal anti-inflammatory drugs considered in the present work.}
  \label{fig:molecs}
  \end{center}
\end{figure}

For the on the fly nonadiabatic dynamics simulations of the molecular systems 
shown in Figure \ref{fig:molecs}, potential energy surfaces (PES) and couplings 
were obtained with the Floating Occupation Molecular Orbitals-Configurations Interaction
(FOMO-CI) method \cite{Granucci2000, Persico2014}, using the
semiempirical AM1 Hamiltonian with standard parameterization. 
We show in Table \ref{tab:casci} the active spaces used in the FOMO-CI
calculations for the systems considered in this work. In all the cases, 
the CI space was of CAS type. 
For all the molecular systems considered, the orbitals included in the FOMO-CI active space
can be classified as $\pi$ and $\pi^*$, at least in the vicinity of the ground state
equilibrium geometry (see Figure S1 of the Supporting Information). For those
systems for which no TD-DFT absorption spectra have been reported in the
literature, i.e. Carprofen and Indomethacin, we have computed the energies of
the first transitions using the TD-CAMB3LYP/6-31+G(d,p) protocol at the ground
state geometries optimized with the same functional and basis set, that will
serve to assess the reliability of the AM1 method in the prediction of the
electronic energies for these systems. The most important
configurations contributing to each electronic state and the orbitals involved
in these configurations can be found in Table S1 and Figures S2 and S3 of the
Supporting Information.
\begin{table}
  \caption{CAS-CI active spaces (electrons/orbitals), number of
    spin-diabatic states (singlets, triplets) and number of trajectories considered in the dynamics 
    simulations. ``Non valid'' trajectories were discarded. ASP$_{(aq)}$ labels aspirin
  in a cluster of water molecules.}
  \label{tab:casci}
  \begin{center}
  \begin{tabular}{lcccc}
    \hline\hline
    molecule     & active     & $S$, $T$     & \multicolumn{2}{c}{\# of trajectories} \\
    \cline{4-5}
                 & space      &              & total          & Non Valid    \\
    \hline
    ASP          & 4/4        &  3, 4        & 401            & 0         \\
    ASP$_{(aq)}$ & 4/4        &  3, 4        & 440            & 0         \\
    IBU          & 4/3        &  3, 2        & 422            & 20        \\
    CAR          & 4/4        &  3, 4        & 203            & 3         \\
    SUP          & 6/5        &  4, 3        & 364            & 60        \\
    IDM          & 6/5        &  3, 3        & 304            & 9         \\
    \hline\hline
  \end{tabular}
  \end{center}
\end{table}

Spin-orbit (SO) couplings, needed for the description of intersystem crossing processes in the dynamics
simulations, were evaluated with a mean field Hamiltonian \cite{Granucci2011}. The relevant
semiempirical parameters for C, O, Cl and N atoms were obtained \cite{Granucci2011} from the splitting
of the ground state atomic terms (for N the ${}^2D^\circ$ term was considered, as the ground state
${}^4S^\circ$ has no splitting) and their respective values are 28.6, 170, 588 and 4 cm$^{-1}$. For
the S atom, we used the same semiempirical spin-orbit parameter (500 cm$^{-1}$) employed in a previous
work on thioguanine \cite{Martinez2014}. 

The molecular dynamics simulations were performed with the fewest switches surface hopping 
method \cite{Persico2014, Tully1990, Granucci2001}. The trajectories were propagated on the
``spin-adiabatic'' potential energy surfaces, obtained diagonalizing the electronic Hamiltonian
(including the SO coupling) in the subspace spanned by the first few spin-defined electronic states
(the ``spin-diabatic'' states, see Table \ref{tab:casci}) \cite{Martinez2014}.  The integration of the
electronic time dependent Schr\"odinger equation was performed with the widely used Local Diabatization (LD)
scheme \cite{Granucci2001}, which is very convenient in the case of
weakly avoided crossings  \cite{Crespo2018} \cite{Plasser2012}. In the present work we used a new
algorithm for the evaluation of the fewest switches transition probability, described in the Appendix,
especially useful
in the case of many weakly interacting states (consider that each triplet state is split in three
components by the SO interaction). Quantum decoherence was approximately taken into account with 
the ODC method \cite{Granucci2010}.
The time step used in the propagation of the trajectories was $\Delta t = 0.1$ fs. 

ASP-solvent interaction has been described within a QM/MM framework with electrostatic
embedding: the aspirin molecule was inserted in a spherical cluster of 775 water molecules, described
with a TIP3P force field. This system will be labeled ASP$_{(aq)}$ from now on.  To avoid evaporation
of water molecules from the surface of the cluster during the dynamics simulations, a constraining
potential was added \cite{Cusati2011}.

The starting conditions for the surface hopping calculations were sampled from thermal equilibrated
trajectories, propagated on the ground state potential energy surface for 50 ps (10 ps
for ASP$_{(aq)}$)
at 300 K with the Bussi-Parrinello thermostat \cite{Bussi2008}. Each surface hopping trajectory starts
with a vertical excitation to a state selected taking into account the radiative transition
probability. The sampling procedure is described in detail in ref. \cite{Persico2014}. The number of
total trajectories obtained for each system is shown in Table \ref{tab:casci}, together with the number of
trajectories discarded for technical reasons that, thus, were excluded from the analysis.  The surface
hopping trajectories were propagated for 10 ps. In discussing the results of the dynamics simulations,
we shall call "state population" to the average population of the spin-diabatic states, evaluated as shown in
ref.\ \cite{Martinez2014}. All the molecular dynamics simulations were carried
out with a developing version of the MOPAC2002 \cite{MOPAC} package, interfaced to the TINKER
\cite{TINKER} Molecular Mechanics package in the QM/MM calculations.

\begin{table}
  \caption{Semiempirical FOMO-CI vertical excitation energies (eV) and oscillator strengths (in
  parenthesis) computed at the ground state equilibrium structure. Experimental data for the
  excitation energies (in terms of maxima of the absorption spectra) are shown in italics.}
  \label{tab:fc}

  \hspace*{-5em}
  \begin{tabular}{lccccc}
    \hline\hline
    transition     & ASP    & IBU  &   CAR  &  SUP  &  IDM   \\
    \hline
    $S_0\to S_1$   & 4.24 \textit{4.13}$^a$ (0.014) & 5.12 \textit{4.56}$^c$ (0.47) & 3.81 \textit{3.76}$^d$ (0.03) & 4.50 \textit{4.29}$^e$ (0.25) & 4.19 \textit{3.87}$^f$ (0.74)  \\
    $S_0\to S_2$   & 4.63 \textit{4.49}$^b$ (0.027) & 5.19 \textit{4.68}$^c$ (0.36) & 3.98 \textit{4.13}$^d$ (0.01) & 4.58 \textit{4.63}$^e$ (0.59) & 4.23 \textit{4.64}$^f$ (0.33)  \\
    $S_0\to S_3$   & 5.87 \textit{5.41}$^b$ (0.76)  & 8.29                   (0.00) & 4.88 \textit{5.17}$^d$ (1.52) & 5.11                   (0.39) & 4.98 \textit{5.39}$^f$ (0.26)  \\
    $S_0\to T_1$   & 3.16         & 3.24        & 3.29        & 2.70        & 2.98         \\
    $S_0\to T_2$   & 3.78         & 3.84        & 3.38        & 2.89        & 3.13         \\
    $S_0\to T_3$   & 3.86         & 8.31        & 3.67        & 3.26        & 3.41         \\
    $S_0\to T_4$   & 4.48         &  -          & 3.97        & 5.81        & 5.59         \\
    \hline\hline
  \end{tabular}
    ${}^a$ Absorption spectrum in DMSO \cite{Govindasamy2014}, shoulder.
    ${}^b$ Absorption spectrum in various solvents \cite{Florey1979,Govindasamy2014,Iwunze2008}.
    ${}^c$ Absorption spectrum in water \cite{Vione2011}.
    ${}^d$ Absorption spectrum in ethanol \cite{Bosca1997}.
    ${}^e$ Absorption spectrum in acetonitrile \cite{Du2014}.
    ${}^f$ Absorption spectrum in water \cite{Ghatak2008,Maity2015}.
\end{table}

\section{Results and discussion}
\subsection{Aspirin (ASP)} 
Our computed semiempirical FOMO-CI vertical
excitation energies for ASP in vacuum, reported in Table \ref{tab:fc}, match quite well the
experimental results, especially if the shoulder at about 300 nm (4.13 eV) of the peak at 276 nm (4.49 eV) in
the spectrum obtained by Govindasamy et al.\ \cite{Govindasamy2014} in DMSO can be identified with the
$S_0 \to S_1$ transition.  Other experimental absorption spectra of ASP recorded in various solvents \cite{Florey1979,Govindasamy2014,Iwunze2008}
exhibit also maxima at 276 nm (4.49 eV) and 229 nm (5.41 eV).
Although red-shifted, our AM1 vertical transitions are also in reasonable agreement with the
B3LYP/6-311G** values reported in \cite{Ye2016}, the second absorption being the most sensitive to the
level of theory employed in the calculations with a deviation amounting to ca. 0.75 eV.  At the
Franck-Condon region, the first two excited
singlets, $S_1$ and $S_2$, both have mixed character $H\to L$, $H-1 \to L$, $H\to L+1$ and $H-1\to
L+1$, with the $H\to L$ configuration having the largest weight for $S_1$, according to the
AM1 FOMO-CI calculations.

\begin{figure}
  \begin{center}
  \includegraphics[width=15.0cm]{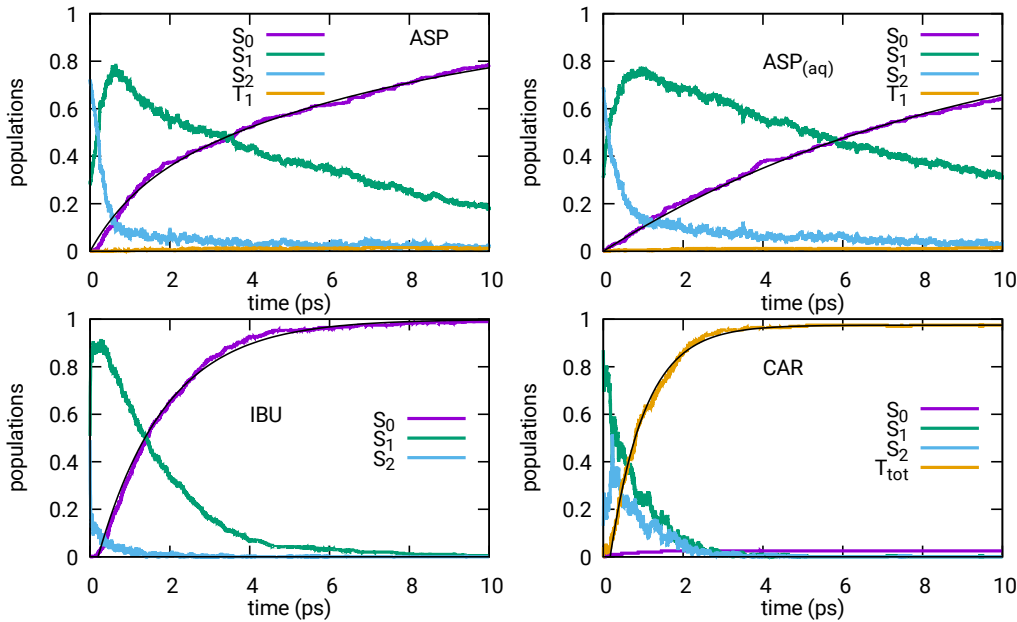}
  \caption{Time evolution of the population of the spin-diabatic states for 
    aspirin (in vacuum and in a cluster of water molecules), ibuprofen and carprofen.
    For carprofen $T_{tot}$ labels the total triplet population. 
    Black curves are fits of the $S_0$ or $T_{tot}$ populations.}
  \label{fig:pop1}
  \end{center}
\end{figure}
In Figure \ref{fig:pop1} we show the time evolution of the population of the spin-diabatic states
of ASP (top left panel). 
Our surface hopping trajectories start with excitations to $S_1$ or $S_2$,
in agreement with the oscillator strengths of Table \ref{tab:fc}. At the beginning, the $S_2$
state is the most populated. The corresponding excitation energy, averaged over the full swarm of 401
trajectories, is 4.49 eV. 
The spin-orbit couplings evaluated at the $S_0$ equilibrium
geometry for the first few electronic states were very weak (of the
order of 1 cm$^{-1}$, or less), for this and the other systems considered in this work.
Within the first picosecond, the $S_2$ population drops from 0.72 to about
0.07, due to $S_2 \to S_1$ transitions. Actually, $S_1$ and $S_2$ are very close in energy, and keep
exchanging population during the 10 ps of dynamics considered: the average number of
$S_2 \to S_1$ and $S_1 \to S_2$ transitions per trajectory 
amounts respectively to 8.5 and 7.8.

\begin{figure}
  \begin{center}
  \includegraphics[width=12.0cm]{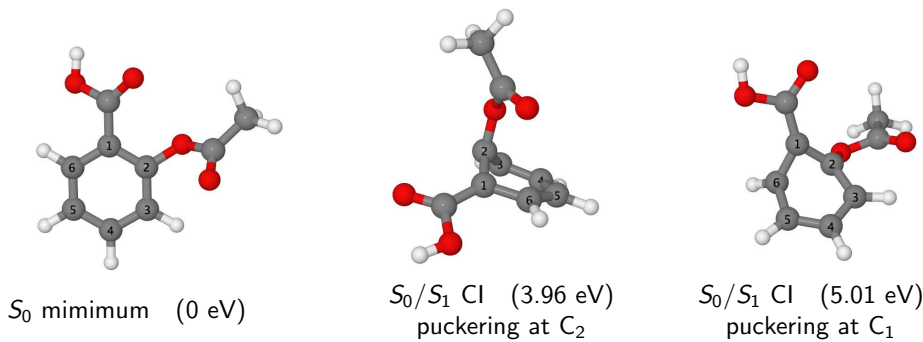}
  \caption{Aspirin. Presented are the AM1 FOMO-CI minimum energy geometries of the ground state, 
    of the $S_1/S_0$ seam with puckering at C$_2$, and of the $S_1/S_0$ seam with puckering
    at C$_1$.}
  \label{fig:aspgeom}
  \end{center}
\end{figure}

Within the first 10 ps, the triplet quantum yield is too low (0.015) to be statistically meaningful: only 6
trajectories out of 401 perform a transition from $S_1$ to $T_1$. Hence, the decay from $S_1$ is almost
exclusively represented by the internal conversion to $S_0$, which happens either through a $S_1/S_0$
conical intersection (40\% of cases), or at large energy difference between the two states (60\%). 
We found in ASP a minimum energy CI (conical intersection) lying at 3.96 eV above the ground state minimum
and corresponding to the puckering of the aromatic ring at the ring carbon atom C$_2$ (see Figure
\ref{fig:aspgeom}), linked to the acetoxy group.  That crossing seam is important in the dynamics. In
fact, all the trajectories making a $S_1\to S_0$ transition with an energy difference between the two
states $E(S_1 -S_0) < 1.5$ eV, show evident puckering of the aromatic ring at C$_2$.  Another
$S_1/S_0$ minimum energy CI, corresponding to puckering of the aromatic ring at C$_1$, is found at
5.01 eV above the $S_0$ minimum, and does not play any role in the dynamics (see Figure
\ref{fig:aspgeom}).

The $S_1$ state of ASP has a ``planar'' minimum (i.e.\ with a planar conformation of the aromatic ring)
at 4.12 eV above the ground state equilibrium energy, easily accessible from the Frank-Condon
region. Puckering of the aromatic ring at C$_2$ would lead to the $S_1/S_0$ CI referred above, which
actually lies 0.16 eV below the planar minimum of $S_1$. However, such a pathway does not seem to be
easily accessed during the dynamics, considering that 60\% of the trajectories in $S_1$ stay trapped
in the planar minimum, eventually hopping to $S_0$ from there, with a large $S_1$-$S_0$ energy
difference (3.97 eV in average).

The rise of the $S_0$ population is well reproduced by the biexponential function
\begin{equation}
  \label{eq:pop}
  P_{S_0}(t) = 1 - w\, e^{-t/\tau_1} -(1-w)\, e^{-t/\tau_2}
\end{equation}
with $w = 0.23$, $\tau_1 = 1.13$ ps and $\tau_2 = 8.21$ ps. At early times (for $t < 2$ ps) the
$S_1 \to S_0$ internal conversion through the $S_1/S_0$ CI dominates with respect to the
transitions from the $S_1$ planar minimum, so that the shorter lifetime $\tau_1$ could be mainly
ascribed to the decay via conical intersection. The situation is reversed at later times.  Notice that
the $S_2$ population drops below 0.1 after just 600 fs, when the population of $S_0$ is still below
0.13.  Therefore, the rise of $P_{S_0}$ is not expected to be much influenced by the $S_2 \to S_1$
decay rate.  It is important to note here that slow decays from weak coupling regions are especially
difficult to be accurately reproduced with surface hopping approaches \cite{Granucci2010}. We expect
such kind of inaccuracies inherent to the method to moderately affect the value estimated for $\tau_2$.

We now switch to ASP$_{(aq)}$ (i.e.\ ASP inserted in a spherical cluster of water molecules). As it
can be appreciated from Figure \ref{fig:pop1}, the decay pattern of ASP$_{(aq)}$ is pretty much
coincident with that of ASP in vacuum, with just a modest slowdown of transition rates, which can be
ascribed to solvent friction. We notice that solvents are quite effective in slowing down the
nonadiabatic transitions. In fact, vibronic coupling is proportional to the nuclear velocities, but
the fastest nuclei, exploring a larger portion of their configurational space, are more subject to the
interaction with solvent molecules. An appreciation of this effect can be obtained by monitoring in
time the average nuclear kinetic energy $\av{E_{kin}}$ of the ASP molecule in different environments.
Comparing $\av{E_{kin}}$ for ASP in vacuum and in the solvent cluster, we note that they start to
diverge after just about 300 fs. By fitting $\av{E_{kin}}$ of ASP in the solvent cluster as in Ref.\
\cite{Cantatore2014} we can also estimate the time constant for the vibrational cooling of ASP in
water solution, which is found to be 30.8 ps.

A closer inspection of the excited states decay dynamics of ASP$_{(aq)}$ reveals a striking difference
with respect to the gas phase results (albeit expected to a certain extent). In particular, 
$S_1 \to S_0$ internal conversion through the $S_1/S_0$ CI is almost completely suppressed. Thus, 
the transitions from the $S_1$ planar minimum represent 93\% of the $S_1 \to S_0$ decay.
This effect might be ascribed to the steric interaction of the chromophore with the solvent cage. In
fact, the puckering of the aromatic ring at C$_2$ leads to a quite important geometrical change for
the whole molecular system and its steric hindrance (as it can be appreciated by looking at Figure
\ref{fig:aspgeom}, the C$_2$-O bond is displaced
from the plane of the ring to a position perpendicular to that plane). Hydrogen bonds with water
molecules may also play a role in this respect, although they are rather established with the COOH
group. In agreement with these findings, the rise of the ground state population can be fitted with a
single exponential function, setting $w=0$ in equation \ref{eq:pop}, and delivering then a value for
$\tau_2 = 9.30$ ps.

\subsection{Ibuprofen (IBU)} 
The experimental UV
spectrum of both neutral and deprotonated IBU in water solution, reported by Vione et 
al.\ \cite{Vione2011}, show a weak band ($\epsilon_{max} \simeq 400$ M$^{-1}$cm$^{-1}$) with a couple of
peaks at 265 nm (4.68 eV) and 272 nm (4.56 eV). Our vertical transition
energies for $S_2$ and $S_1$ (see Table \ref{tab:fc}) are in reasonable
agreement with the positions of the experimental peaks and calculated
absorptions \cite{Musa2007}, both being about 0.5 eV too high compared to the
experimental results, but 0.1-0.4 eV too low with respect to the calculated
values, respectively. In this case, the $S_1$ and $T_1$ states 
have $H\to L$ character, while $S_2$ and $T_2$ have $H -1 \to L$ character (see Figure S1). 

At the beginning of the dynamics $S_1$ and $S_2$ are populated, with an average
excitation energy of 5.07 eV. The time evolution of the spin diabatic
populations is qualitatively similar to ASP, and can be schematized as $S_2
\rightleftharpoons S_1 \to S_0$. However, both the decay from $S_2$ to $S_1$
and from $S_1$ to $S_0$ are faster for IBU. Concerning the $S_2 \to S_1$
internal conversion process, the $S_2$ ``planar'' minimum (i.e.\ with a planar
conformation of the aromatic ring, thus easily accessible from the
Franck-Condon region) corresponds to a $S_2/S_1$ CI, at variance with ASP,
which explains the ultrafast decay of the $S_2$ state for IBU. 

\begin{figure}
  \begin{center}
  \includegraphics[width=12.0cm]{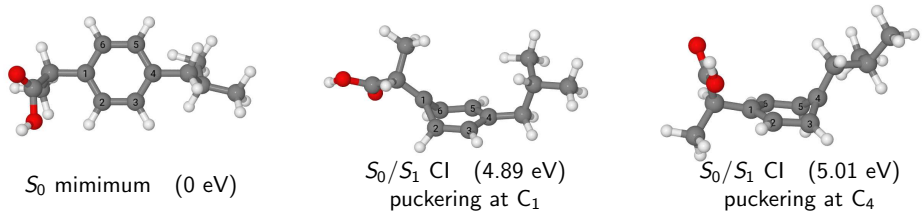}
  \caption{Ibuprofen. Presented are the AM1 FOMO-CI minimum energy geometries of the ground state, 
    of the $S_1/S_0$ seam with puckering at C$_1$, and of the $S_1/S_0$ seam with puckering
    at C$_4$.}
  \label{fig:ibugeom}
  \end{center}
\end{figure}
The deactivation of $S_1$ towards $S_0$ involves several CI regions, that correspond
to geometries where the aromatic ring is puckered. In particular, the most important $S_1/S_0$ conical
intersections involve the puckering of the ring at the carbon atoms C$_1$ and C$_4$ (see
Figure \ref{fig:ibugeom}), and will be labeled in the following CI$_p$ and
CI$_a$. These degeneracy points are, respectively, located 4.89 eV and 5.01 eV
above the $S_0$ minimum. Very close to these intersections, we found a $S_1$
planar minimum which lies 4.97 eV above the $S_0$ equilibrium geometry.
About 55\% (30\%) of the $S_1\to S_0$ transitions go through the CI$_a$
(CI$_p$). The rest is conveyed through conical intersection regions which
require the puckering at other carbon atoms of the ring, lying slightly higher
in energy, in the range 5.03--5.15 eV.  At variance with ASP, no slow decay
from the $S_1$ planar minimum is found for IBU. As a result, the rise of the
$S_0$ population can be fitted by a single exponential (see Figure
\ref{fig:pop1}), with a time constant of 1.89 ps.

It is known that IBU undergoes photolysis after UVB or UVC excitation in solution, for both the
anionic deprotonated and the neutral forms \cite{Vione2011,Yuan2009}. In a TD-DFT study, Musa and
Eriksson \cite{Musa2007} showed that the T$_1$ state of deprotonated IBU has a very small barrier
versus decarboxylation (only 0.3 Kcal/mol). This is in agreement with our FOMO-CI calculations, where
the T$_1$ state of deprotonated IBU is found to be unstable with respect to decarboxylation.
Vione et al.\ \cite{Vione2011} found quite different photolysis quantum yields for neutral and
deprotonated IBU (respectively, 1.0 and 0.3) suggesting that the neutral and deprotonated species would probably undergo dissociation via two different mechanisms. Our nonadiabatic dynamics simulations indicate that the
photolysis of neutral IBU would occur from the hot ground state.

\subsection{Carprofen (CAR)} 
At the ground state equilibrium geometry, the three rings of CAR have an almost planar conformation,
with a very modest pyramidalization at the N atom, which disappears at $S_1$ and $S_2$ minima and
becomes more pronounced at the $T_1$ equilibrium geometry. The $S_1$, $S_2$ and $T_1$ minima lie
respectively at 3.71, 3.91 and 2.90 eV above the ground state equilibrium energy. Consistently with
the much larger difference between adiabatic (i.e.\ minimum-minimum) and vertical transition energies
(see Table \ref{tab:fc}), $T_1$ shows a larger geometrical distortion compared to $S_1$ and $S_2$ from
the Franck-Condon point. 
The experimental absorption spectrum \cite{Bosca1997} of CAR in ethanol shows a peak at
300 nm (4.13 eV) with a broad side band extending up to 350 nm, and a more intense band at about 240 nm
(5.17 eV).  In the gas phase, TD-CAMB3LYP predict the first three transitions, all of $\pi\to \pi^*$
character, in the range between 4.3 and 5.3 eV, i.e 4.29, 4.70 and 5.30 eV, See Table S1 and Figure
S2 of the Supporting Information. Our computed FOMO-CI vertical transitions 
$S_0\to S_1$,  $S_0\to S_2$, and $S_0\to S_3$ 
are found at slightly lower energies with respect to the experimental data, see table \ref{tab:fc}.
The fluorescence spectrum of CAR in ethanol \cite{Bosca1997} shows two peaks at 367 nm (3.38 eV) and
352 nm (3.52), which may be compared with our FOMO-CI vertical transition energies $S_1 \to S_0$ (3.63
eV) and $S_2 \to S_0$ (3.84 eV). The phosphorescence spectrum \cite{Bosca1997,Moser2000} has a maximum
at about 430 nm (2.88 eV); the corresponding calculated $T_1 \to S_0$ vertical transition energy is
2.46 eV. Overall, the AM1 FOMO-CI energies show a reasonable agreement with the experimental data
available.

In the simulations of CAR excited state dynamics, the starting populations of the $S_1$ and $S_2$
states are 0.13 and 0.87, respectively, with an average excitation energy of 3.82 eV.  The $S_3$
state, lying much higher in energy, was not included in the molecular dynamics simulations.  Following
excitation, the $S_1$ and $S_2$ population is transferred very rapidly to the triplet states
$T_1$-$T_4$, which are very close in energy. The total triplet population rises with a time constant
of 1.04 ps, and the triplet quantum yield after 10 ps is 0.975. All the triplets considered in the
simulations receive population from the singlets, with about 80\% of the transitions going to $T_3$
and $T_4$, the closest to $S_1$ and $S_2$. Once on the triplet manifold, the trajectories hop to $T_1$,
so that the total triplet population is always almost coincident with that of $T_1$. 

According to the experimental results of Bosca et al.\ \cite{Bosca1997}, the triplet quantum yield of
CAR in ethanol solution after excitation at 355 nm is 0.37, and the fluorescence quantum yield is
small but not negligible (0.06). Although our simulations correctly account for the decay to the
triplet manifold, the population transfer to the triplets appears to be severely overestimated.  It
seems very unlikely that this can be attributed to an overestimation of the SO couplings, considering
that our calculated values for the latter amount to 1 cm$^{-1}$ or less. These discrepancies with the
experimental results could be rather ascribed to inaccuracies in the description of the PES. However,
the solvent may also play a role in slowing down the transition rate to triplet states.

\subsection{Suprofen (SUP)} 
The experimental absorption spectrum of SUP in acetonitrile \cite{Du2014} (where the molecule is
mainly present in its neutral form) shows two peaks at 264 nm (4.63 eV) and 289 nm (4.29 eV), which are
red shifted upon deprotonation \cite{Du2014,Musa2009}. 
Comparing these data with the calculated $S_0\to S_1$ and $S_0\to S_2$ vertical transition energies of
Table \ref{tab:fc}, the $S_1$ and $S_2$ states appear to be too close in energy, in between the
experimental peaks, still not far from them. By comparison with the experiment, the energy of the
$S_3$ state is probably underestimated by our calculations.  Our AM1 FOMO-CI calculations predict the
first two $\pi\to \pi^*$ transitions in the same region of the spectrum as the
B3LYP/6-31G(d,p) protocol \cite{Musa2009}.

\begin{figure}
  \begin{center}
  \includegraphics[width=12.0cm]{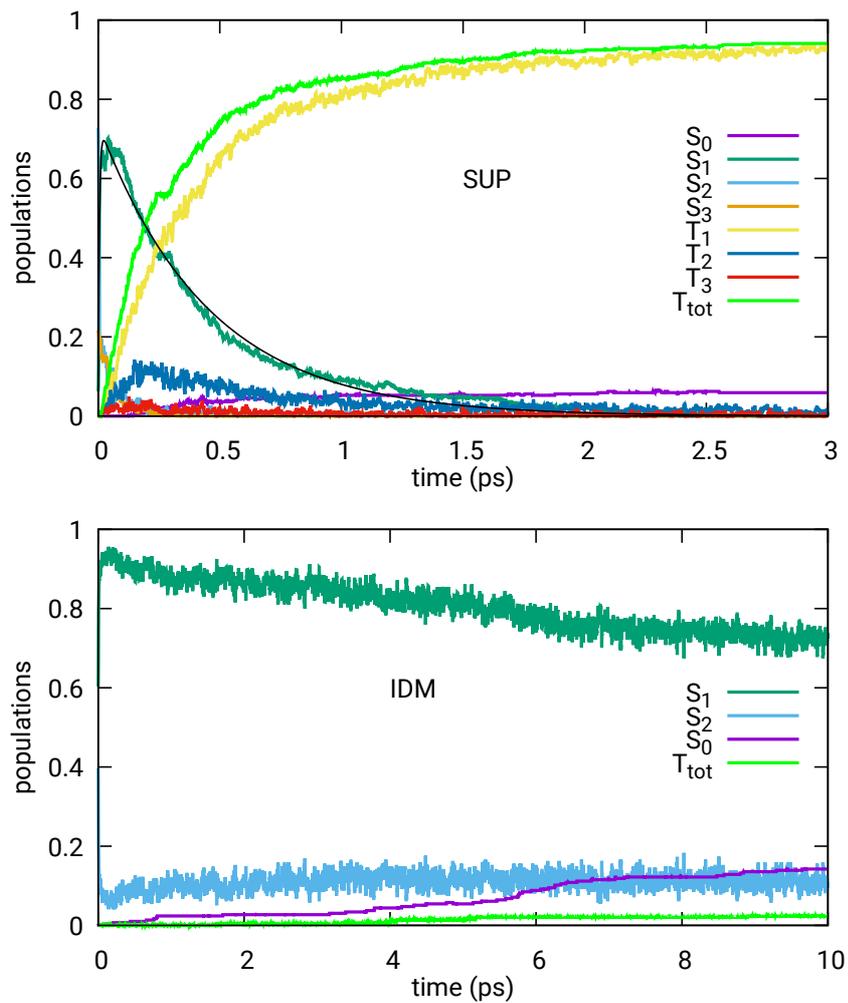}
  \caption{Time evolution of the populations of the spin-diabatic states for 
    suprofen (upper panel) and indomethacin (lower panel).  Note the different time scale for the two
    molecules (for suprofen, state populations are stable in the range 3--10 ps). 
    $T_{tot}$ is the total triplet population.
    The black curve is a fit of the suprofen $S_1$ population.}
  \label{fig:pop2}
  \end{center}
\end{figure}
\begin{figure}
  \begin{center}
  \includegraphics[width=12.0cm]{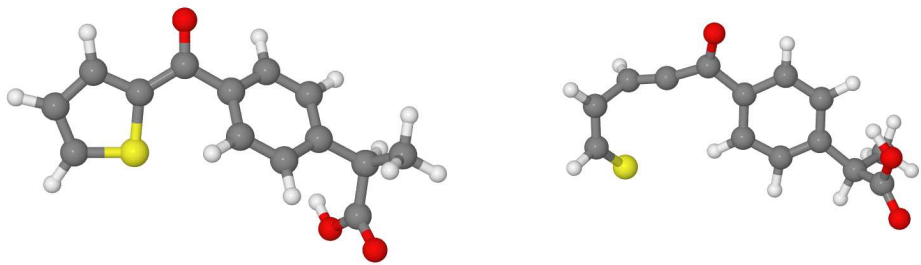}
  \caption{Ground state minimum of suprofen (left), and an exemplary
    geometry at which $S_1\to T_2$ and $S_1 \to S_0$ transitions takes place (right).}
  \label{fig:supgeom}
  \end{center}
\end{figure}

The surface hopping trajectories are started from $S_1$, $S_2$ and $S_3$ (initial populations 0.06,
0.73 and 0.21, respectively), with an average excitation energy of 4.74 eV. Trajectories in the $S_3$
state decay rapidly to $S_2$, and even more rapid is the transition from $S_2$ to $S_1$ (see Figure
\ref{fig:pop2}). In fact the $S_2$ minimum, readily accessible from the Franck-Condon region,
coincides with a $S_2/S_1$ CI.  From $S_1$ we have mainly transitions to $T_2$ and $T_3$, that follow
two distinct pathways.  In particular, $S_1\to T_2$ transitions 
occur mainly in a degeneracy region where the bond between the sulfur and the carbon
atom directly linked to the carbonyl group is broken, or at least is very stretched (see Figure
\ref{fig:supgeom}). In fact, the
average S-C distance at $S_1$-$T_2$ hops is R(S-C)=2.70 \AA.
In all the trajectories but two (the total number of valid trajectories for SUP amount to 304, recall Table
\ref{tab:casci}), the S-C bond is readily reconstructed after the hop. The transitions to $T_3$ occur
instead in $S_1/T_3$ near degeneracy regions, where the thiophene ring is distorted (for example
through the puckering at the S atom) although not opened. Once in the triplet manifold, very rapidly
the $T_1$ state is reached, as the three triplets lie very close in energy to each other (recall the
energies of the triplets at the Franck Condon region collected in Table \ref{tab:fc}).

The decay pattern reported above is in agreement with the character of the excited states.  In fact,
for $S_1$ and $T_2$ the excitation involves occupied orbitals mainly located on the thiophene moiety
(the most important configurations for both states are $H\to L$ and $H\to L+1$, see Figure S1), while
for $T_3$, which has a mixed character $H-2\to L$ and $H-2\to L+1$, the phenyl ring is mainly
concerned. Accordingly, both $S_1$ and $T_2$ show a larger equilibrium S-C distance than $S_0$
(R(S-C) = 1.75, 1.69 and 1.68 \AA, respectively). 

A minor but not negligible number of trajectories undergo $S_1\to S_0$ decay. After 10 ps the $S_0$
population is 0.06, and the triplet quantum yield amounts to 0.94. Again, the $S_1$-$S_0$ transitions
occur at large S-C distances, where $S_0$, $S_1$, $T_1$ and $T_2$ are almost degenerate (see Figure
\ref{fig:supgeom}).
The $S_1$ population can be fitted by a biexponential function, appropriate for a two-step first
order irreversible decay kinetics
\begin{equation}
  P_{S_1} = w (e^{-t/\tau_2} - e^{-t/\tau_1}) \; .
  \label{eq:pops1}
\end{equation}
We obtain $w=0.74$, $\tau_1=0.005$ ps and $\tau_2 = 0.45$ ps. In their transient absorption
experiments with SUP in acetonitrile, Du et al. \cite{Du2014} registered a lifetime of 2.3
ps for $S_1$. Moreover, in correspondence with the decay of the signal attributed to $S_1$, the rise
of an absorption attributed to $T_1$ was apparent. This is in nice agreement with our results,
although our lifetime $\tau_2$ for the $S_1$ decay is about five times smaller than the experimental
value.  Large amplitude motions, involved in the transitions from $S_1$ to triplets (for example, the
opening and reforming of the thiophene cycle) are expected to be hindered in solution, slowing down
the $S_1$ decay.  The solvent may then explain, at least in part, the difference between our gas phase
lifetimes and those obtained in the transient absorption experiments in solution.

\subsection{Indomethacin (IDM)} 
The experimental absorption spectrum of IDM in water \cite{Maity2015,Ghatak2008} shows two peaks at
320 nm (3.87 eV) and at 267 nm (4.64 eV), to be compared with the vertical excitation energies $S_0\to
S_1$ and $S_0\to S_2$  at 4.19 and 4.23 eV, shown in Table \ref{tab:fc}. The AM1 energies of these two
absorptions are only 0.1 eV below the vertical transitions predicted by the TD-CAMB3LYP/6-31+G(d,p)
protocol, compare Table \ref{tab:fc} and Table S1 of the Supporting Information. The fluorescence and
phosphorescence spectra are
also reported \cite{Maity2015,Dabestani1993}, with maxima respectively at 380 nm (3.26 eV, in water)
and 475 nm (2.61 eV, in various organic solvents). The corresponding calculated $S_1\to S_0$ and
$T_1\to S_0$ vertical transition energies are found at 3.52 eV and 1.99 eV, respectively. Both $S_1$
and $S_2$ show some charge transfer character from the indole moiety to the benzoyl group. In fact,
the Mulliken charge of the indole ring increases by about 0.15  (0.18) going from $S_0$
to $S_1$ ($S_0$ to $S_2$). 

At the beginning of the nonadiabatic dynamics simulations for IDM the population of $S_1$ and $S_2$ is
0.6 and 0.4, respectively, with an average excitation energy of 4.15 eV. Due to the very small
$S_2$-$S_1$ energy difference at the Franck-Condon geometry (only 0.04 eV), most of the population of
$S_2$ is transferred to $S_1$ within the first 50 fs, which is by far the most populated state during
the first 10 ps that last our simulations. Notice, however, that the trajectories keep hopping back and forth
between $S_1$ and $S_2$, as it may be appreciated from the noisy character of $P_{S_2}$ and $P_{S_1}$
curves in Figure \ref{fig:pop2}. This is due to the fact that the $S_2$-$S_1$ gap stays small at the
geometries visited during the dynamics. In particular, the $S_2$ state is only 0.5 eV 
above $S_1$ at the $S_1$ equilibrium geometry, while the $S_2$ minimum coincides with a $S_2/S_1$ CI.
The average energy difference $S_2$-$S_1$ oscillates around about 0.5 eV during the dynamics, 
which is an indication of the fact that most of the trajectories wander around the $S_1$ minimum region.
It is from that region that the transitions $S_1\to S_0$ occur, leading to a $S_0$ population of
0.14 after 10 ps. A fit of the decay of $P_{S_1} + P_{S_2}$ delivers a lifetime of 55 ps. 
As already stated above, the accurate description of this kind of decay represents
a particularly difficult task for surface hopping. Moreover, our simulation is limited to the very
beginning of the IDM excited state dynamics (see Figure \ref{fig:pop2}). For these reasons, 
our estimation of 55 ps for the singlet lifetime has to be taken as a rough approximation of the real value.

At the $S_1$ minimum, the energy difference $S_1$-$T_3$ and $S_1$-$T_2$ is only 0.11 and 0.35 eV,
respectively.  A small number of trajectories (7) undergo intersystem crossing from $S_1$ to the
nearby triplets $T_2$ or $T_3$, leading to a triplet population of 0.02 after 10 ps. 
It is of course not possible to extract a triplet quantum yield from these data. However,
they are in agreement with the experimental observation of  phosphorescence for
IDM.

\section{Concluding remarks} 
We have undertaken a computational investigation of the excited state dynamics of some NSAIDs, using
the surface hopping approach and including the spin-orbit interaction to account for intersystem
crossing processes. Although PES and wavefunctions were obtained in a semiempirical framework
retaining the standard AM1 parameters, overall our FOMO-CI energies showed a quite good agreement with
the experimental spectroscopic data available.

ASP and IBU, which share the benzene group as chromophore, show a relatively fast decay to the ground
state, with very little or no population of the triplet states, with only a modest solvent slowdown
for ASP$_{aq}$. The decay is especially fast in IBU,
which has a singlet lifetime of only $\sim 2$ ps. Transitions to $S_0$ are triggered by distortions of
the aromatic ring (especially for IBU), in particular involving the carbon atoms linked to the
substituents. 

CAR and SUP show ultrafast decay to $T_1$, with time constants for the rise of the
triplet population of 1.04 and 0.31 ps, respectively, and triplet quantum yields close to 1.
However, the decay pattern is quite different: for CAR, $S_1$ and $S_2$ are very close in energy to
$T_3$ and $T_4$, both at the Franck-Condon point and at the $S_1$ and $S_2$ minima, and the
singlet/triplet transitions take place without relevant distortions of the polycyclic conjugated
system. For SUP, transitions to triplets involve severe deformations of the thiophene moiety.

An experimental time-resolved absorption study of excited state dynamics of SUP in acetonitrile by Du
et al.\ is available in the literature \cite{Du2014}. They obtain a decrease of the $S_1$ signal
simultaneous to the increase of the $T_1$ signal, which is only in qualitative agreement with our
results, as their time constants for the singlet decay is about five times larger than ours (2.3
versus 0.45 ps, respectively).  However, it is our opinion that the solvent could play a role in
slowing down the intersystem crossing rate, especially considering the large amplitude motion we found
involved in this process for suprofen.

A still different behavior is shown by IDM, where the $S_1$ state show a very slow decay, with a
dynamics which is far from being completed after the first 10 ps. Among the NSAIDs considered in the
present work, clearly the IDM dynamics is the least suitable for a surface hopping investigation.

The molecular dynamics results from this study, together with the information
inferred from the static works of other groups on this family of NSAIDS, reveal
the complexity of the new descriptors to be designed to capture the stability
or reactivity of drugs exposed to light of different wavelengths and the
deficiency of simple models such as for instance the one based on the magnitude
of HOMO-LUMO gap. 
In fact, the traditional HOMO-LUMO gap model fails to predict the photoreactivity/photostability of 3
out of the 5 systems considered in this work by classifying IDM (8.15 eV), CAR (8.14 eV), and SUP (8.81
eV) as unlikely phototoxic systems attending to their H-L gaps which exceed the threshold value of 8.1
eV \cite{Peukert2011} (H-L gaps provided within parenthesis correspond to the most stable conformers
calculated at AM1 level of theory).

This and other works bring forward the
importance of the topography of the potential energy surface far away of the
Franck-Condon region, revealing crucial information on the inherent stability
of these chromophores connected to the existence of excited state minima that
could potentially trap the population in the excited state or the accessibility
of decaying funnels or for the transfer of population to manifolds of other
multiplicities. Interestingly, in the case where several decay routes are
competitive, time resolved approaches allow determining their relative
importance in the global decay mechanisms. Finally, our study also discloses
the importance of accounting for solvent solute interactions, which as shown
here might importantly tune the decay mechanisms and lifetimes. 

\section*{Acknowledgments}
N. A. thanks the Marie Curie Actions, within the Innovative Training Network-European Joint Doctorate
in Theoretical Chemistry and Computational Modelling TCCM-ITN-EJD-642294, for financial support.
G.G. acknowledges funding from the University of Pisa, PRA\_2017\_28. 
I.C. gratefully acknowledges the ``Ram\'on y Cajal'' program and financial support from the Project
CTQ2015-63997-C2 of the Ministerio de Econom\'{\i}a y Competitividad of Spain.

\appendix
\section{} 
Transition probabilities in surface hopping are usually obtained according to Tully's ``fewest
switches'' prescription \cite{Tully1990,HammesSchiffer1994}.  It has been noted \cite{Bajo2014} that
the fewest switches algorithm used in local diabatization (LD) schemes
\cite{Granucci2001,Crespo2018,Mai2018} is physically not well grounded and subject to numerical
instabilities, expecially in the presence of many weakly interacting states.  In view of
these difficulties, we propose here an alternative procedure.

For a given surface hopping trajectory, the electronic wavefunction at time $t$ is expanded on the
adiabatic basis $\kket{\bm{\varphi}}$
\begin{equation}
  \Psi_{el}(t)  = \sum_j C_j(t) \kket{\varphi_j(t)} 
\end{equation}
The population of state $\kket{\varphi_j}$ is then $P_j(t) = |C_j(t)|^2$.
In the LD scheme, the integration of the time-dependent Schr\"odinger equation
for the electrons is performed in an alternative ``locally diabatic'' electronic basis
$\kket{\bm{\eta}}$, spanning the same subspace as the adiabatic basis
$\kket{\bm{\varphi}}$, and defined so as to be (approximately) constant in the integration time
step $\Delta t$
\begin{align}
  \kket{\bm{\eta}(0)} & = \kket{\bm{\varphi}(0)}  \\
  \kket{\bm{\eta}(\Delta t)} & \equiv \kket{\bm{\eta}(0)} \simeq
   \kket{\bm{\varphi}(\Delta t)} \bT^\dagger
\end{align}
Here we set $t=0$ at the beginning of the time step. The unitary matrix $\bT(\Delta t)$ is obtained
by L\"owdin orthonormalization of the overlap matrix $\ovl{\bm{\varphi}(0)}{\bm{\varphi}(\Delta t)}$
between the adiabatic wavefunctions at the beginning and at the end of the time step.

Let $\kket{\varphi_k}$ be the current state (i.e.\ the state on which PES the trajectory is running). 
We label $W_k$ the total transition probability from state $k$. In agreement with the fewest switches
algorithm, $W_k$ is given by
\begin{equation}
  W_k = \max \left\{ 0, \; \frac{P_k(0) -P_k(\Delta t)}{P_k(0)} \right\}
\end{equation}
The problem is how to partition $W_k$ among the adiabatic states $j \neq k$ in order to find the
transition probabilities $\tkj$ from the current state $k$ to the other adiabatic states $j$.
Notice that in a two state case there is nothing to partition, the transition probability $\tkj$
being just equal to $W_k$. We assume then in the following that the number of states is larger than 2.

For $W_k=0$ we set $\tkj=0$. Assuming $W_k >0$, we want to define $\tkj >0$ such that $\sum_{j\neq k}
\tkj =W_k$.
To this aim, we notice that $\tkj$, according to the fewest switches algorithm, 
has to depend on the population increment of state $j$ within the
time step $\Delta t$ and on the coupling between states $k$ and $j$, which is in turn proportional to
the overlap $\ovl{\varphi_k(0)}{\varphi_j(\Delta t)}$. We introduce then
the following quantities
\begin{equation}
  \label{defsk}
  x_{kj} = \frac{|\bT_{kj}| \sqrt{\Delta P_j}}{S_k} \qquad \mathrm{with }\quad 
        S_k=\sum_{j\neq k} |\bT_{kj}| \sqrt{\Delta P_j}
\end{equation}
where $\Delta P_j = \max \{ 0, \; P_j(\Delta t) -P_j(0) \}$ is the population increment for state $j$.
The non negative quantities $x_{kj}$ define the partition, so that the fewest switches transition
probability is given by
\begin{equation}
  \tkj = x_{kj} W_k \; .
\end{equation}
The peculiar functional form of equation \ref{defsk} for $S_k$ has been chosen considering that,
according to the LD propagation algorithm of ref.\ \cite{Granucci2001}, the $P_j(\Delta t)$
probabilities are proportional to the square of the $\bT$ matrix elements.  
A good indication of the fact that the partition induced by the factors $x_{kj}$ is correct would be
that the normalization term $S_k$ is equal to the decrement of the current state population. Actually,
this can be easily shown for a system that wanders in a region of weak coupling, where $P_k(0) \simeq
1$ and $P_j(0) \simeq 0$ for $j\neq k$ (as it would be enforced by any algorithm aimed at introducing
quantum decoherence effects).  Therefore, according to equation 16 of ref.\ \cite{Granucci2001},
$P_j(\Delta t) \simeq |\bT_{kj}|^2$. Moreover,  $W_k P_k(0) \simeq W_k \simeq 1-|\bT_{kk}|$. Hence, as
$\bT$ is unitary, we have $S_k \approx W_k P_k(0)$, which is what we wanted to prove.

Good properties of the algorithm outlined above are: (i) easily implemented and based
on quantities already available in a LD computation; (ii) the partition proposed is physically sound
and not subject to numerical instabilities; (iii) the sum of the $\tkj$ is equal to $W_k$ by
construction (at variance with a previously proposed improvement \cite{Bajo2014} of the original fewest
switches algorithm \cite{Granucci2001} for LD). 


\bibliography{biblio}

\end{document}